# Inter-slit Coupling in Gold Film Hole Arrays

Itai Carmeli,[1#] Roman Walther,[2#] Reinhard Schneider,[2] Dagmar Gerthsen,[2,3] Yaron Kaufman,[4] Ayala Shvarzman,[1] Shachar Richter,[1] Hagai Cohen[5*]

1. School of Chemistry and Center for Nanoscience and Nanotechnology, Tel-Aviv University, Tel Aviv 69978, Israel
2. Laboratory for Electron Microscopy, Karlsruhe Institute of Technology (KIT), D-76128 Karlsruhe, Germany
3. Center for Functional Nanostructures, Karlsruhe Institute of Technology (KIT), D-76128 Karlsruhe, Germany
4. Faculty of Materials Engineering, Technion – Israel Institute of Technology, Haifa 32000, Israel
5. Department of Chemical Research Support, Weizmann Institute of Science, Rehovot 76100, Israel

**#** - equal contribution
**\*** - corresponding author

## Abstract

Inter-slit interactions across one-dimensional arrays of sub-µm rectangular holes in gold films are explored. Using electron energy loss spectroscopy combined with scanning transmission electron microscopy, a series of cavity standing waves is resolved, indicating particularly high interslit interactions, about an order of magnitude larger than the intra-slit edge to edge coupling. Pronounced signal enhancements are thus induced, dominated by short-range interactions and high mode-localization, while yet, relatively long-range coherence is retained. The sub-nm electron beam, in spite of principal differences from broad-area probes, yields results similar to extraordinary optical transmission (EOT). Implications to EOT mechanisms, including its sub-wavelength, off-resonance limit, are pointed out.

The discovery of extraordinary optical transmission (EOT) through thin metal films[1] has drawn attention to the role of the light induced activity at the mask itself. Thanks to extensive studies[2-7], understanding of the complex EOT mechanism has been greatly improved and possible applications were proposed.[8-10] EOT is believed to involve efficient coupling of the illuminating beam to surface plasmon polariton (SPP) modes propagating on the metal film; light-SPP coupling that is greatly enhanced under the introduction of holes with sharp edges.[10-12] A single hole can thus manifest EOT. However, it is widely accepted that periodic arrays of holes propose improved photon-SPP coupling, assisted by lattice-associated momentum transfers during the scattering process. The range of interactions between resonances localized in different slits becomes, in this respect, an interesting question, particularly in view of the dominant role played by *short* range interactions in part of the current EOT literature. [12, 13]

Attempts to optically map the near electromagnetic (EM) field above and around the holes generally suffer from limited resolution.[14, 15] Complementarily, electron energy loss spectroscopy (EELS) in scanning transmission electron microscopy (STEM) was applied to study isolated holes in free-standing gold films at sub-nm transversal resolution.[16,17] By providing effective coupling to both far-field (FF) and near-field (NF) EM components,[18, 19] the focused electron beam revealed tight hybridization between SPPs and vacuum-supported wave-guide (WG) resonances of the single slit. In contrast to energy-filtered TEM (EFTEM) images,[20, 21] the scanned focused beam is a unique, needle-like antenna, complementary to the broad-beam illumination techniques. As shown below, this fact becomes particularly important in studies of inter-slit interactions. Here, STEM-EELS is exploited to investigate the local field enhancements in one-dimensional (1D) arrays of rectangular holes with sub-µm size. A dramatic effect of nearest neighbors is found, combined however with relatively long-range interference between the edge-confined modes.

Experiments were performed on FEI Titan microscopes [22] at 300 keV electron energy, as detailed in the Supplementary Information. All samples discussed here consist of rectangular 900x180 nm$^2$ holes in a 1D array, with pitch (*p*, inter-slit distance) values of

280, 450, 900 nm, including a reference of an isolated hole representing pitch→∞. Key improvements in the quantification of low-energy, sub-eV, signals were achieved by applying the Richardson-Lucy algorithm (RL),[23, 24] with a vacuum reference spectrum as an input for the point spread function (PSF) and with a limitedly small number of iterations. Subsequent background subtraction with smooth bi-exponent functions allowed further improvements in the evaluation of peak intensities. All intensities presented here are normalized with respect to the zero loss peak height.

Fig. 1a presents a series of EEL spectra along a line-scan across two slits of a four-slit array, as indicated in the inset. Below the Au surface plasmon at 2.4 eV, cavity resonances of the L=900 nm slit length are well observed. EELS detection of similar resonances was already reported,[16] enabled by the breakdown of momentum conservation and the appearance of scattering channels with momentum transfer $\neq \omega/v$, where $\omega$ is the SPP frequency and v is the beam velocity.[24-27] Here, the fundamental line at ~0.5 eV (a value depending on both electron beam position and array parameters) is red-shifted considerably, compared to the bare EM standing wave of this slit length.[2, 10] Fig. 1b shows EEL spectra taken at different positions (indicated by colored crosses) along the 2$^{nd}$ slit of a 60-slit array. It demonstrates how the fundamental line multiples undergo intensity variations that match 'classical' standing waves: $\omega_2$ vanishes at the center, L/2, and reaches maximum intensity at L/4; $\omega_3$ exhibits reduced intensity at L/4 and $\omega_4$ vanishes at both places. The energy dependence on multiple order (n) is nearly linear, as expected at the light-line branch of SPPs. Importantly, none of these multiple modes tends to vanish toward the slit corner, as may be naively expected.

The spectra in Fig. 1a exhibit clear near-field (NF) characteristics: sharp decay tails associated with SPP localization around the slit wall.[28] Yet, on top of the NF character in Fig. 1a, the signal strength at the outer wall is drastically weaker than at its neighboring wall and, in the inner slit, an additional increase is seen. Fig. 2a presents the intensity profile of the fundamental line $\omega_1$ across an eight-slit array. Except for the array ends, it demonstrates small variations only, superimposed on the wall-specific NF tails. Similar profiles across larger arrays (not shown) indicate oscillations, slightly above the experimental error level, that depend on the pitch value (*p*).

Line intensity enhancements of $\omega_1$ across the very first slit in the array are plotted in Fig. 2b (squares) as a function of the number of slits (N). A striking change is realized under the introduction of a second slit, but then, the dependence on N is rapidly reduced. Analysis of the peak heights (circles) – as opposed to integrated intensities of the fundamental line - yields consistently larger enhancements, which points to line-width effects. Complementarily, details of the signal enhancements in a two-slit system are shown in Figs. 3a,b for different interslit distances. The outer wall spectra, Fig. 3a, taken at positions marked by red dots in the STEM images, are not affected much by the distance between neighboring slits and, in fact, are all rather similar to the single slit spectrum. In contrast, marked spectral differences are seen in Fig. 3b, near the inner wall (yellow dots in the STEM images).

All these results point to high interslit coupling, much stronger (about an order of magnitude) than the intra-slit edge-to-edge coupling.[29] The former is attributed to mediating SPPs, excited on the top ($SPP^{top}$) and bottom ($SPP^{bot}$) faces of the gold film, and interacting with SPPs on the slit walls ($SPP^{wall}$) across common slit edges. At a first glance, the coupling across metal strips seems to decay with the strip width, on a scale of ~ 200-300 nm, as may be expected for modes confined to a given edge and decaying away toward the next edge. A closer inspection of the spectra in Fig. 3b reveals, however, nonmonotonic behavior due to coherence effects: The third multiple signal of the 900 nm pitch, at $\omega_3 \approx 1.5$ eV, is 'surprisingly' stronger than (1) $\omega_1$ in the same spectrum and (2) $\omega_3$ of the 450 nm pitch. Both results correlate with the phase difference across the separating metal strip, given by $2\pi d/\lambda$, where d is the strip width and $\lambda$ is the inspected SPP wavelength. It becomes close to $\pi$ for $\omega_1$ and $\omega_3$ in the 900 nm and 450 nm case, respectively (d/$\lambda$= 270/600 and 720/1800).[30] As expected, none of these latter two effects shows up at the outer wall, Fig. 3a.

In close similarity with Fig. 2a, calculated optical transmission intensity profiles (at the EM resonance, Fig. 4 in Ref. 13)[13] suggest sharp signal suppressions near the two array ends, followed by (depending on N) small variations only across the array central part. Here too, the major edge effect is followed by similar mid-array features: (1) maximum intensity is observed at the center of our N=4 array, Fig. 1a, and, in contrast (note the N-dependence in Fig. 4 of Ref. 13), a slight decrease toward the center shows up in the longer, N=8 system (Fig. 2a); (2) small intensity oscillations seem to characterize the large N arrays (not shown). Moreover, quantitative agreement between Fig. 2b and the EOT maxima calculated in Fig. 3 of Ref. 12 is noted.

Interestingly, however, in the corresponding EOT literature these features arise under broad-beam illumination, subject to interference between the incident beam and the diffracted SPP modes at *neighboring* slits.[12] The present experiment is principally different: As evidenced from the dominant NF character of the inspected signals, direct e-beam interaction with distant slits is weak.[31] It couples strongly to its nearest edge (mainly through SPP$^{wall}$) and, indirectly only, via interslit EM interactions, to more distant slits. Our results reflect, therefore, a different interference mechanism.

A complementing intriguing feature arises in Fig. 3: The inner wall spectra, Fig. 3b, undergo significant line *narrowing* in contrast to the common expectation for *broadening* of coupled localized modes. Accordingly, peak-height enhancements >4 (the theoretical limit for two identical coupled modes) are obtained in Fig. 2b. This seeming inconsistency suggests that collective modes of the *metal strip as a whole* emerge, with oscillator strengths that exceed those of the isolated edge. Note also that the intensity decrease in Fig. 2b (~15% in peak intensity) starts for N>4 and is realized (although less pronounced) across slits 4-5 in Fig. 2a. Since destructive interference in Fig. 2 corresponds to ~3 periods (3x280 ~ 900), these effects correspond to anti-phase across three periods of *strips*, pointing to a dominant role of strip activity.

Importantly, the electron beam couples preferentially to the wall-SPPs. Note specifically in Fig. 3, outer vs inner-wall spectra, how *enhancements* are 'switched on' upon

*terminating* (by a neighboring slit) the free propagation of top/bottom modes.[32] A relatively long-range phase coherence, retained under the confined excitation source, is a notable feature of the system, enabled by the top and bottom-SPPs and expected to play a significant role in EOT as well. To better explain our description, Fig. 4a illustrates how the top and wall-SPPs share similar field components. Thus, with corresponding WG components, no significant distortions are needed to achieve continuity of fields across common slit edges, in full agreement with the polarization selectivity observed in EOT.[7] To a certain extent, a single-slit description based on linear combinations of these modes can therefore give insight to involved field-mapping calculations and become a useful means for understanding the effect of neighboring slits. Our data point, however, to limits of this description[33] and, in particular, to distortions taking place at the long tail of $SPP^{top}$ ($SPP^{bot}$), far above (below) the film surface, see tail illustration in Fig. 4b. For a strip width of $d=\lambda/2$, one realizes antiphase oscillations at its edges and the corresponding top and bottom SPP tails along z should be strongly distorted in a destructive manner. In contrast, for a narrow strip, $d \ll \lambda$, all four strip faces oscillate in phase, with an extended field above and below the film. Thus, the electron-beam sensitivity to such collective strip effects, in spite of its high xy localization, is enhanced along path sections relatively far from the film plane.

Further consequences regarding the EOT mechanism should be noted. First, it is broadly claimed that the array effect in optical studies originates mainly in the improvement of beam-film (i.e. to SPPs) coupling via momentum transfer to the lattice. The combined data above show that the dielectric collective response within the film is, by itself, greatly enhanced for sufficiently close slits.

Second, while mode confinement to slit edges is retained, our data suggest a shift in EM activity from the inner-slit space (for a single slit) to *strip*-supported domains (under the introduction of a neighbor); an effect predicted also for the EOT configuration.[13]

Third, our data demonstrate how cavity boundary conditions relax on (and above) the metal strips: signals in Fig. 1b do not fully vanish towards the narrow wall of the slit (the cavity ends).[34] As symmetry with respect to the center, L/2, is preserved, light interaction via SPPs of longer, off-resonance, wavelengths can be realized.

Finally, exploiting the superior EELS mapping capabilities, we were able to reveal the interplay between high localization of the field and its extended realization. This duality would help the focusing of energy under *broad area* illumination through the holes. The close match in SPP field components, $E_x$ and $E_z$, across a slit edge, see Fig. 4a, gives rise to improved top-to-bottom coupling of film faces via mediating wall-SPPs and, hence, to an antenna-like energy collection.[35]

In summary, investigating array effects with a focused electron beam allows a unique view on film dielectric response. We reveal particularly strong inter-slit coupling, much stronger than the intra-slit interactions, combined with pronounced SPP amplitude enhancements, while confinement to slit edges is retained and, intriguingly, extended coherence shows up. This unique combination of array properties is proposed to play a key role in sub-wavelength, off-resonance EOT.

**Acknowledgements**: We thank Tsofar Maniv for helpful comments. This research was supported by the Weizmann's Helen and Martin Kimmel Center for Nanoscale Science.
.

29. Estimations are based on nearesr neighbors couplings, to be elaborated in a future report.
30. It is d and not p (pitch) that explains the Fig. 3 data.
31. e-beam coupling strength a few percents of the major term is estimated (at the relevant beam position) for residual contributions from neighboring slit edges.
32. This effect also clarifies that the beam couples mainly to wall-SPPs other than top/bottom modes.
33. Both in a single slit and in arrays we find the decay tails sharper than expected naively, indicating high k-values involved, probably contributed (in addition to other sources) by corresponding rim edges of the slit.
34. The red-shift in the fundamental frequency, as well as the intra-slit coupling mechanism, are partially attributed to those strip-supported longer wavelength SPPs.
35. The EOT effect means collection of energy from a broad area and transferring it to the other side of the film (an antenna-like mechanism). Our results show how neighboring slits can help it: while the area per slit goes down for lower pitch values, the activity jumps up via hybridization of strip SPPs. Thus, wall-SPPs serve as energy condensers from top face to bottom.

**Figure captions**

**Fig. 1**: (a) EEL spectra taken along a line-scan across half of a 4-slit array, skipping over the metal strip as indicated. Inner slit spectra are grey marked; (b) Representative EEL spectra along the wall of the second slit in a 60-slit array, showing the fundamental line, $\omega_1$, and its first multiples ($\omega_{2-4}$) below the bare surface plasmon, $\omega_s$ (hints for higher multiples are also indicated). Corresponding STEM images are given as insets, with three illustrated standing wave multiples, n=1-3. All spectra shown are after 7 iterations with the RL algorithm, followed by a bi-exponent background subtraction. Pitch in both samples is 280 nm.

**Fig. 2**: (a) The intensity profile of the fundamental line $\omega_1$ across an 8-slit array, pitch=280 nm, as derived from the area under the $\omega_1$ line; (b) intensity enhancements in arrays of different number of slits, represented by inner/outer ratios, as measured at a 20 nm distance from the inner and outer walls of the *first* slit: area under the $\omega_1$ line (squares) and $\omega_1$ peak height (circles). Note the emergence of intensity *decrease* across n=4,5 in (a) and N>4 in (b), where n is the slit number within a given array.

**Fig. 3**: The effect of inter-slit distance in a 2-slit system, as expressed by EEL spectra recorded at a fixed, 20 nm distance from the (a) outer and (b) inner wall. Corresponding STEM images are shown. Red and yellow dots mark the positions spectra were taken at. The scale bar represents 200 nm.

**Fig. 4**: **Illustration of field components for the y-cavity resonances:** (a) Both wall-SPPs propagating down/up in z-direction and top-SPPs propagating along x from one slit to the other consist of $E_x$ and $E_z$ field components. The focused electron-beam (indicated by a solid arrow) couples preferentially to $E_z$ components and further requires $k_z$ momentum transfer (the latter available by wall SPPs and at slit edges). A broad-area light illumination is indicated by dashed arrows to illustrate why, under x-polarization, it gains improved coupling to both SPP modes. Note that y-cavity boundaries strictly hold for wall-SPPs only, not for the top-SPPs. Note also that the phase at a neighboring wall (a red cosine indicated) varies with the pitch value. (b) Top and bottom-SPPs consist of a long z-tail into the vacuum, on a scale of the SPP wavelength, $\lambda$. When a neighboring slit is introduced, distortions are expected mainly at the tail, subject to the phase across the metal strip.

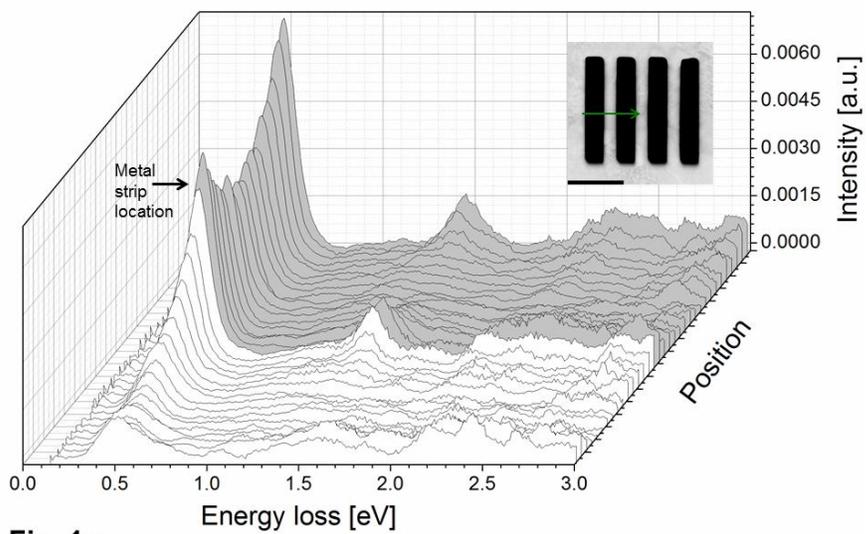

**Fig. 1a**

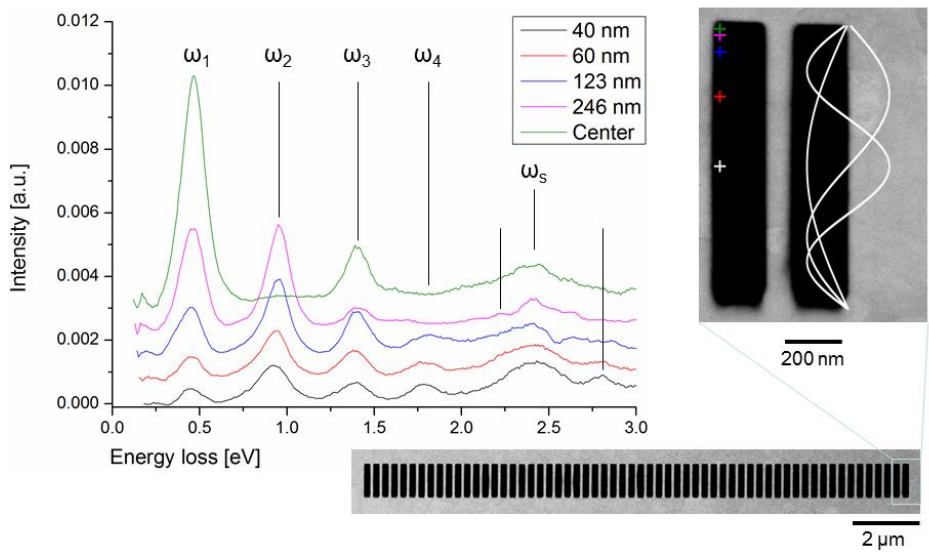

**Fig. 1b**

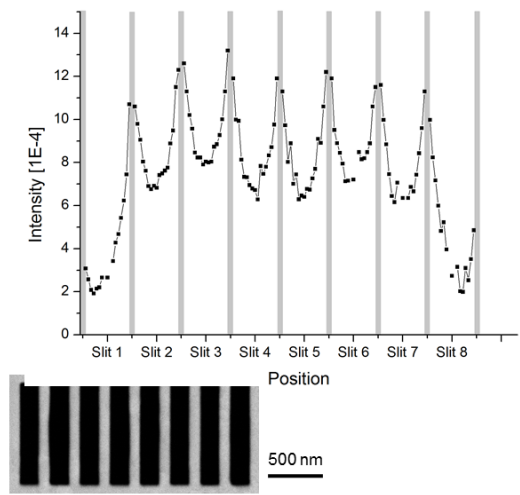

**Fig. 2a**

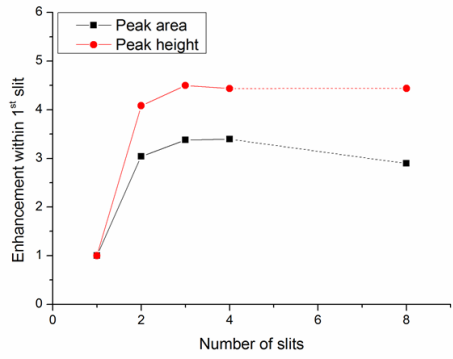

**Fig. 2b**

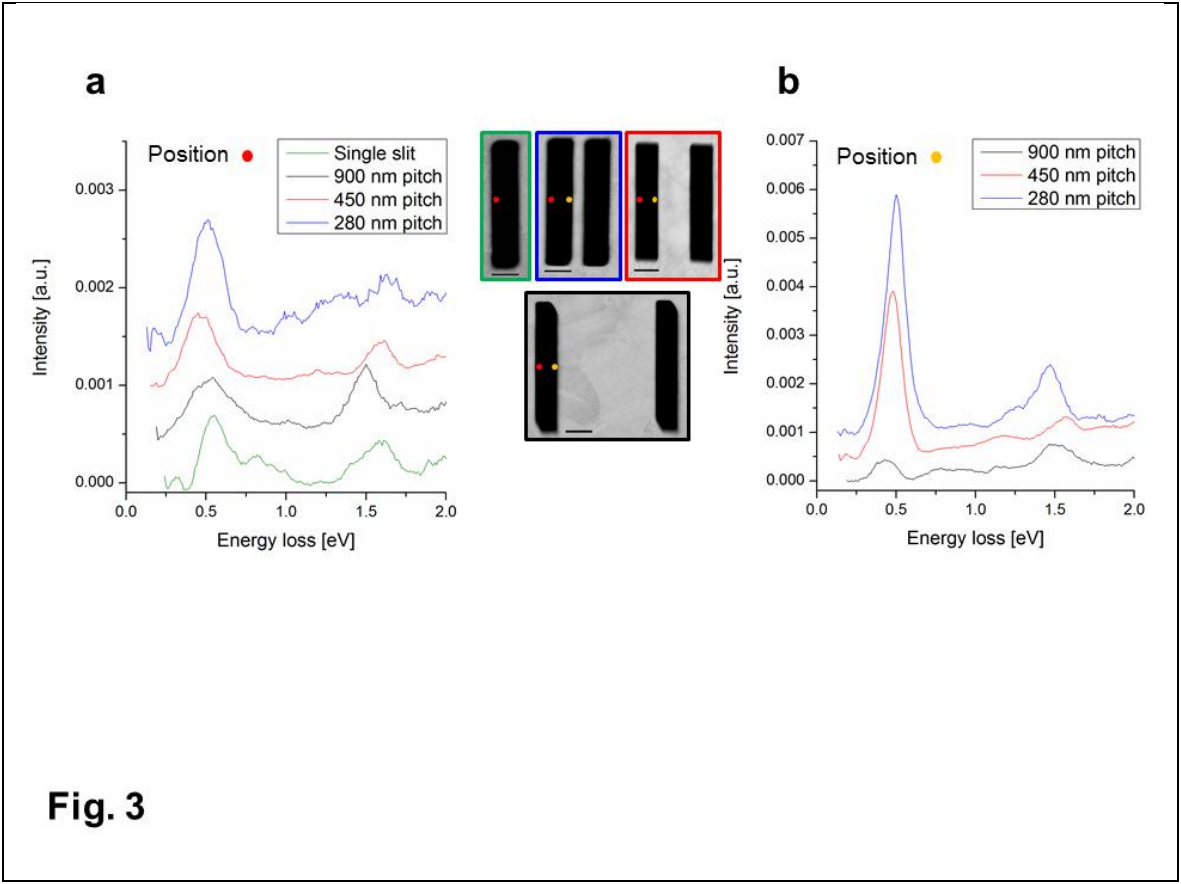

Fig. 3

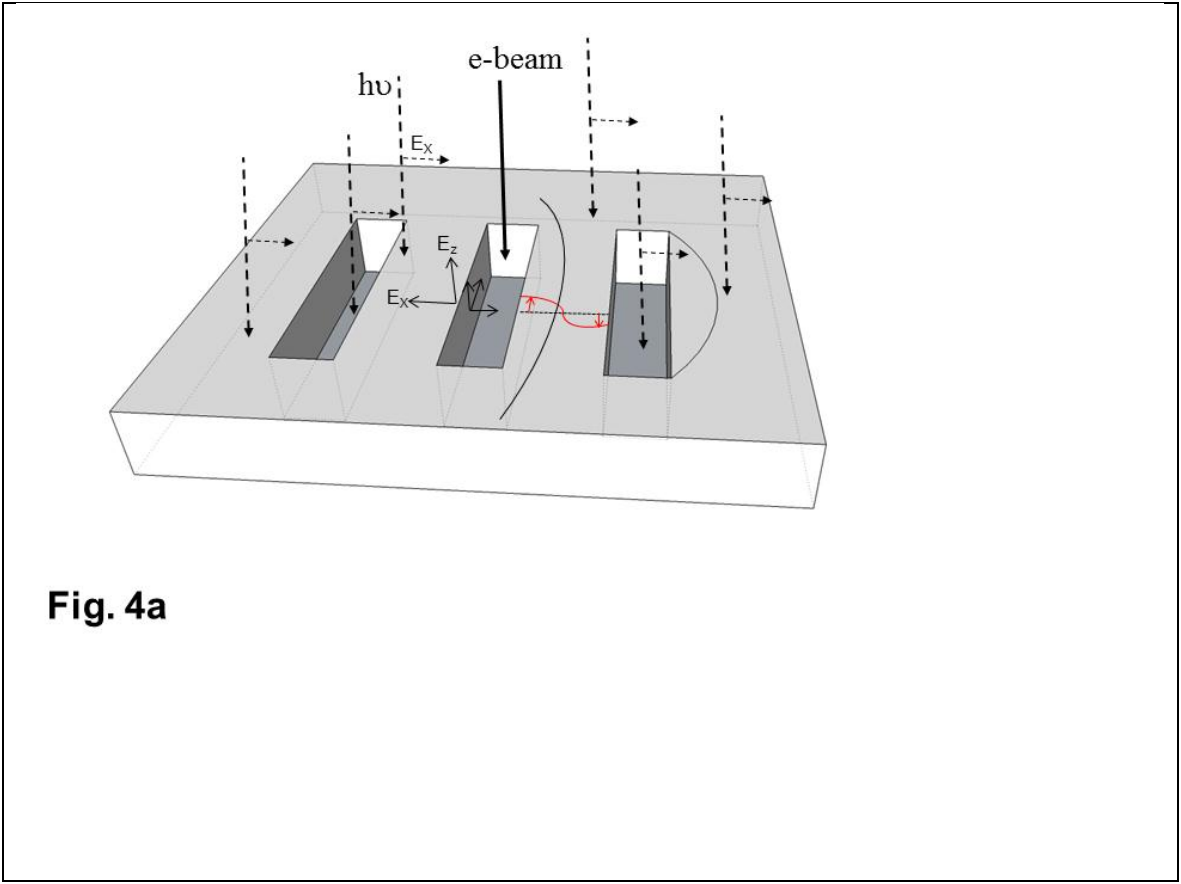
Fig. 4a

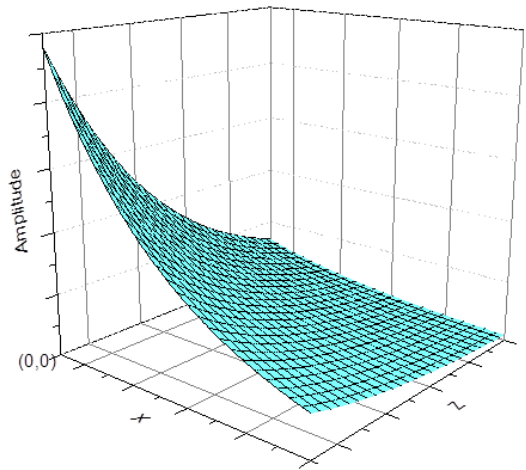

**Fig. 4b**